\documentclass[a4paper, 11pt]{article}
\usepackage[margin=1in]{geometry}
\usepackage[english]{babel}
\usepackage{amsmath, amssymb, amsfonts}
\usepackage{multirow, multicol, booktabs}
\usepackage{mathrsfs, mathtools, slashed}
\usepackage{microtype}
\usepackage{hepunits}
\usepackage{csquotes}
\usepackage{titling} 
\usepackage{authblk, abstract}
\usepackage{blindtext}
\usepackage[dvipsnames, table]{xcolor}
\usepackage[pdftex,
            pdfauthor={M. Althaf, Triparno Bandyopadhyay},
            pdftitle={Potential for the discovery of the protophobic
            boson at the STCF},
            pdfsubject={Feasibility study of the Super tau-charm
            facility (STCF) for the detection of the protophobic 17 MeV
        boson, hypothesised as a solution to the ATOMKI nuclear
    anomalies, using TrackEff and Madgraph.},
            pdfkeywords={ATOMKI anomaly, Protophobic boson, Super
            tau-charm facility, TrackEff},
            pdfproducer={Latex with hyperref},
            pdfcreator={pdflatex},
            colorlinks=true,
            allcolors=CadetBlue
]{hyperref}

\usepackage{orcidlink}
\usepackage[
       backend=biber,
       style=numeric-comp,
       sorting=none
   ]{biblatex}
\hypersetup{%
pdftitle={Ruling out the protphobic boson},
pdfsubject={Your PDF subject},
pdfauthor={Triparno Bandyopadhyay, M. Althaf},
pdfkeywords={BES-III, protophobic, displaced vertex}
}
\usepackage[nameinlink]{cleveref}

\newcommand{\refone}[1]{\textcolor{black}{#1}}
\newcommand{\reftwo}[1]{\textcolor{black}{#1}}
\newcommand{\reftwot}[1]{\textcolor{black}{#1}}

\title{{Potential for the discovery of the 
protophobic boson at the STCF}}

\author[1]{Althaf M.\,\orcidlink{0009-0002-7170-2870}\thanks{\href{emailto:am7636@srmist.edu.in}{\texttt{althaphysics@gmail.com}}}}
\author[1]{Triparno Bandyopadhyay\,\orcidlink{0000-0002-6744-8972}\thanks{\href{emailto:gondogolegogol@gmail.com}{\texttt{gondogolegogol@gmail.com}}}}
\affil[1]{\small Department of Physics and Nanotechnology, SRM Institute of Science and Technology,
    SRM Nagar, Kattankulathur, Tamil Nadu 603203.
}

\begin{document}
\maketitle {\textbf{Abstract:} We study the morphology of the main drift
chamber (MDC) to be built around the collision point of the proposed
Super tau-charm facility (STCF), to check \reftwo{its suitability for
discovering the 17 MeV protophobic boson} (X17 boson), hypothesised as a
solution to the persistent ATOMKI nuclear-transition anomalies. Using
the TrackEff framework, we perform detector-level simulations of the
STCF MDC, and evaluate displaced-vertex sensitivities towards the
protophobic boson, across the relevant mass-coupling parameter space. We
study benchmark scenarios with visible and dark decay channels to
perform likelihood-based significance estimates in order to determine
the \reftwo{5~$\sigma$} discovery reach for the protophobic boson. We
find that STCF can \reftwo{potentially} discover the protophobic boson while tolerating $\sim 10^4$ background events for specific regions of the parameter space. Our analysis establishes the first feasibility study of displaced light-boson searches at the STCF, motivating a full Geant-4 simulation. }


\section{Introduction}%
\label{sec:Introduction}

There are intriguing observations pertaining to the transitions of light
nuclei which are at tension with the standard model (SM) of particle
physics. These observations, corresponding to the E1, M0, and M1 nuclear
transitions of elements like \(^8\)Be,\(^{16}\)O, \(^4\)He, \(^{12}\)C
have most recently been reported by the group at the ATOMKI facility in
Debrecen,
Hungary~\cite{Vitez:2008zz,Krasznahorkay:2015iga,Krasznahorkay:2017bwh,Krasznahorkay:2021joi,Krasznahorkay:2022pxs,Krasznahorky:2024adr}.
However, the anomalies persist from the middle of the
1990s~\cite{deBoer:1996qdk,deBoer:1996qdk,deBoer:1997mr,deBoer:2001sjo,Krasznahorkay:2006zf}.
These different experiments report enhanced correlation between final
state \(e^+\)--\(e^-\) scattering angles, which cannot be explained by
internal or external pair conversions. \reftwo{To ameliorate the
    tensions, a protophobic boson (\(X\)-boson), of mass \(\simeq\)
    \SI{17}{MeV}, has been proposed as a promising new physics (NP)
model~\cite{Feng:2016ysn, Feng:2020mbt}} (Also check, Refs.
\cite{Fayet:1986rh, Kozaczuk:2016nma, Barducci:2022lqd}). In this work, we
attempt  to \reftwo{study} the detection possibilities of the
protophobic boson at the proposed Super \(\tau\)-Charm facility
(STCF)~\cite{Achasov:2023gey,Bao:2025aic}. 

The STCF is a proposed \(e^+\)--\(e^-\) collider with a centre of mass
(CoM) energy ranging from \SI{3}{GeV} to \SI{7}{GeV}, to be constructed
in China, as an update to the BEPCII/BESIII experiment. The STCF is
expected to produce a peak luminosity of \(5\times 10^{34}\)
cm\(^{-2}\)s\(^{-1}\) and an integrated luminosity of
\SI{15}{ab}\(^{-1}\) -- \SI{20}{ab}\(^{-1}\) (\SI{1}{ab}\(^{-1}\) per
year)~\cite{Achasov:2023gey,Bao:2025aic}. The interaction point (IP) is
expected to be surrounded by a drift chamber---the main drift chamber
(MDC)---in the barrel region. We propose to \reftwo{exploit} the
tracking capabilities of this MDC to look for displaced vertex (DV)
signatures produced by the protophobic boson. We focus on the DV
signatures as our interest lies in the intermediate coupling ranges,
\(\sim 10^{-3}\) to \(\sim 10^{-6}\), a range over which the DV strategy
works better than prompt (see, e.g., \cite{Fayet:1980rr}) or invisible
searches~\cite{Bertholet:2021hjl,Ferber:2022ewf,Bandyopadhyay:2022klg}. 

In modelling the MDC, we use the framework
\textsf{TrackEff}\footnote{\href{https://zenodo.org/records/13831999}{https://zenodo.org/records/13831999},
CC 4.0}~\cite{Bertholet:2025lcr} which has been published
recently. \textsf{TrackEff} is essentially a \textsf{Python} module which allows
us to model the detector geometry and the magnetic field for
tracking detectors. The package takes in as input the location of the DV
and the momenta of the final state charged tracks (\(e^-e^+\) for our
study). It then propagates the particle tracks in the magnetic field,
computing the number of hits produced by them in the detector, using the
detector geometry defined by us. In our simulations, we construct the
STCF MDC geometry in \textsf{TrackEff} and define our criteria for a
track to be `accepted'. 

Our end goal is to compute the number of events per CoM energy per
\(X\)-boson mass and coupling, given the known luminosity associated
with each CoM energy. We focus on a small window around \SI{17}{MeV},
where the \(X\)-boson is hypothesised to reside. For each parameter
point, we compute the number of background events admissible for a
\reftwo{5~\(\sigma\)} significance discovery. We thus establish a baseline
feasibility for the detection prospects of the \(X\)-boson at the STCF\@.
Hence, we extend the physics goals of the STCF to solving one of the
remaining unexplained phenomena in particle physics. 

It goes without saying that a full detector-level validation, using
Geant-4~\cite{GEANT4:2002zbu}, is required to substantiate the claims
about \reftwo{\(5~\sigma\)} discovery. However, as mentioned above, our current
aim is to establish a feasibility baseline using the \textsf{TrackEff}
model. We note, in the lack of an actual detector, it is not possible to
validate \textsf{TrackEff} against real STCF data. However,
\textsf{TrackEff} has been cross-verified against real Belle II data and
shows \reftwo{consistency} within 10\% with data~\cite{Bertholet:2025lcr}. Given
that, our results are expected to be conservative, at least at the
order-of-magnitude level. 

The paper is divided as follows, in \cref{sec:ATOMKI} we list and describe
the different anomalies recorded in the nuclear transitions by the
ATOMKI group and by other collaborations, in \cref{sec:models} we
discuss our parametrisation of the protophobic model, then in
\cref{sec:STCF}, we describe the STCF main drift chamber and how we
model it in our code, in \cref{sec:results} we report our results,
before concluding in \cref{sec:conclusion}.


\section{\reftwo{Anomalies in \texorpdfstring{\(e^+e^-\)}{e+e-} pair creation in nuclear decays}}%
\label{sec:ATOMKI}
Anomalous results in light nuclear transitions have persisted since the
mid 1990s. We have given a snapshot of all these reports in
\cref{tab:anomaly}. The first report was by the group at the Institut
f\"{u}r Kernphysik, University of
Frankfurt~\cite{deBoer:1996qdk,deBoer:1997mr,deBoer:2001sjo}. The group
reported a deviation from expectations in the angular correlation of the
final state electron-positron pairs~\cite{Rose:1949zz, Rose:1963zz,
PhysRev.133.B1368} in the M1 \(e^+e^-\) decays of the \SI{17.6}{MeV}
level (\(J^\pi=1^+, T=1\)) of\(~^8\)Be. Taking into account both
internal pair conversion (IPC) and external pair conversions
(EPC)~\cite{Borsellino:1953zz, Hart:1959zza, Olsen:1963zz} does not
explain the deviation. The same discrepancy was not observed in the E1
decays of the 17.2 MeV level (\(J^\pi=1^-, T=1\)) of \(~^{12}\)C. The
deviation in the M1 decays had a significance of \reftwo{4.5~\(\sigma\)}. The
anomaly was observed for large values of the correlation angle
\(\omega\). Both EPC and IPC fall off drastically for large separation
angles~\cite{Rose:1949zz}, more so for low \(Z\) nuclei, hence, this
anomaly was attributed to an exotic boson which was supposed to have
been produced in the nuclear reaction. In the rest frame of the boson,
the \(e^+e^-\) pair comes out back to back. In practice, the angle is
smaller due to the Lorentz boost. A fit by the group predicted the mass of
the exotic boson to be equal to \SI{9}{MeV}.
\begin{table}[htpb]
    \centering
    \caption{A list of all the anomalies reported in nuclear transitions
    which could be explained by an exotic boson of mass in the MeV
ranges.}
    \label{tab:anomaly}
    \begin{tabular}{l l l}
        \toprule
        References & Transitioning Nucleus & Exotic Mass (MeV)\\
        \midrule
         Ref. \cite{deBoer:1996qdk,deBoer:1997mr,deBoer:2001sjo} & M1 decays of the \SI{17.6}{MeV} and \SI{14.6}{MeV}
               levels of \textsuperscript{8}Be & {9.0} \\
               Ref. \cite{Krasznahorkay:2006zf} & M0 transition in
               the \SI{10.95}{MeV} \(0^-\to0^+\) decay of \(~^{16}\)O &
               9.0 -- 10.0\\
         Ref. \cite{Vitez:2008zz} & M1 decay of the \SI{17.6}{MeV}
                level of \textsuperscript{8}Be & \(12.0\pm2.5\)  \\
         Ref. \cite{Krasznahorkay:2015iga,Krasznahorkay:2017bwh} & M1 decays of the \SI{17.6}{MeV} and \SI{18.15}{MeV}
                levels of \textsuperscript{8}Be & \(16.70\pm0.35\pm0.50\) \\
         Ref. \cite{Krasznahorkay:2021joi}   & \textsuperscript{3}H(\(p,\gamma\))\textsuperscript{4}He reaction 
            & \(16.94\pm0.12\pm0.21\)\\
        Ref. \cite{Krasznahorkay:2022pxs} & E1 decay of the
        \SI{17.2}{MeV} level of \textsuperscript{12}C& \(17.03\pm0.11\pm0.20\)\\
        Ref. \cite{Anh:2024req} & M1 decay of the \SI{17.6}{MeV}
                level of \textsuperscript{8}Be & \(16.66\pm0.47\pm0.35\)  \\
        Ref. \cite{Krasznahorky:2024adr} & Giant Dipole Resonance decay of
        \textsuperscript{8}Be& \(16.95\pm0.48\pm0.35\)\\
       \bottomrule
    \end{tabular}
\end{table}

A similar deviation in the M1 decay of~\(^8\)Be was observed when the
same experiment was repeated at the \reftwo{1 MV} Van de Graaff accelerator of
ATOMKI~\cite{Vitez:2008zz} in Debrecen, Hungary. The correlation in the
angular separation could be explained by fitting \(e^+e^-\) pairs
originating from the decay of a \(12.0\pm2.5\)~MeV exotic boson. The
same group also reported a similar excess in the magnetic monopole, M0,
transition of the \SI{10.95}{MeV} \(0^-\to 0^+\) transition of
\(~^{16}\)O~\cite{Krasznahorkay:2006zf}. This discrepancy \reftwo{could
be} explained by a boson of mass between 9.0 and \SI{10.0}{MeV}. 

A follow up experiment was performed at the \reftwo{\SI{5}{MV}} Van de Graaff
accelerator at
ATOMKI~\cite{Krasznahorkay:2015iga,Krasznahorkay:2017bwh}. The first of
these experiments focussed on the same M1 decay of~\(^8\)Be. This
experiment did not find any substantial excess in the \SI{17.6}{MeV}
isovector transition but did find a discrepancy in the \SI{18.5}{MeV}
isoscalar transition at \reftwo{6.8~\(\sigma\)}. The anomaly
\reftwo{could be} explained by
invoking an exotic boson with a fitted mass of \(16.70\pm0.35\pm0.50\)
MeV. 

\reftwo{Apart from the}\(~^8\)Be system, such an anomaly was found
in\(~^3\)H\((p,\gamma)^4\)He reactions as
well~\cite{Krasznahorkay:2021joi}, in an experiment performed at the 2
MV Tandetron accelerator at ATOMKI\@. The peak in the angular
correlation spectrum at a large angle, 115\(^\circ\), \reftwo{could be}
explained by the creation and the subsequent decay of a light particle
during the proton capture process to the ground state of the\(~^4\)He
nucleus. The significance of the discrepancy varied between
\reftwo{6.6~\(\sigma\) to \(8.9~\sigma\)} for varying proton beam
energies. The averaged value of the fitted mass of the exotic was found
to be \(16.94\pm0.12\pm0.21\) MeV. 

Further support for the existence of the exotic boson came from the E1 decay of
the \SI{17.2}{MeV} level of\(~^{12}\)C. The correlation in the angular
distribution, around 155--160\(^\circ\), was found in the E1 \reftwo{to} ground state
decay of the \SI{17.2}{MeV} \(1^-\to0^+\) transition of\(~^{12}\)C.
\reftwo{The
mass of the corresponding 
exotic~\cite{Feng:2020mbt}---which could explain the anomaly---}was fitted to be
\(17.03\pm0.11\pm0.20\) MeV. For different beam energies, the confidence
for the fitted mass ranged from \reftwo{3~\(\sigma\) to 8~\(\sigma\)}.

The results at the ATOMKI accelerators were cross-checked at the 5SDH-2
Pelletron accelerator at the VNU University of Science at Hanoi,
Vietnam~\cite{Anh:2024req}. The same\(~^8\)Be system was studied and the
same excess in the angular separation of the \(e^+e^-\) was found at
around \(135^\circ\), at a significance greater than \reftwo{\(4~
\sigma\)}. The fitted mass of the exotic boson that could explain the
excess was found to be \(16.66\pm0.47\pm0.35\) MeV.

\begin{figure}[t]
    \centering
    \includegraphics[scale=1]{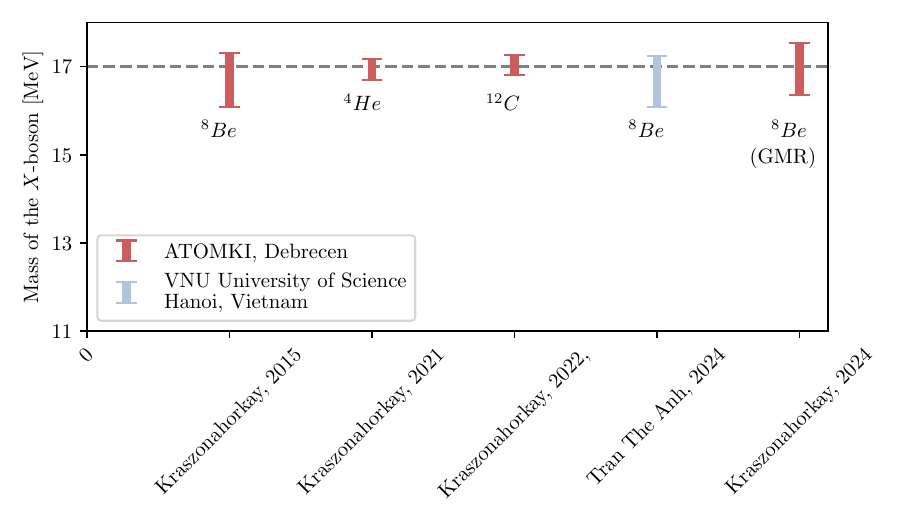}
    \caption{\reftwot{A compilation of the different results 
        {indicating the possible existence of a new \SI{17}{MeV} boson.} The
        different experiments are indicated by the year of publication
and the nucleus undergoing transition.}}
    \label{fig:exp}
\end{figure}

To further test the vector nature of the exotic boson couplings, the
decay of the giant dipole resonance of\(~^8\)Be was studied, again at
the \SI{2}{MV} Tandetron accelerator of ATOMKI, Debrecen. The fit of the
theoretical prediction of the E1+M1+exotic to the experimental data
gives a \reftwo{\(>10~\sigma\)} confidence for the exotic. The measured
invariant mass of the exotic was found to be
\(16.94\pm0.48\pm0.35\)~MeV. \reftwot{We have plotted the different
experimental results, that can potentially be explained by the existence
of a \SI{17}{MeV} boson, in \cref{fig:exp}. }

The results of all these experiments \reftwo{may strongly indicate the
presence of an exotic boson with vectorial couplings and a mass around
17~MeV.} 

A series of experiments, planned or running, have also been
\reftwo{devised to probe} the existence of the exotic boson. These
include nuclear collision experiments at JINR,
Dubna~\cite{Abraamyan:2023hed}, at the Mu3e
experiment~\cite{Perrevoort:2023jry}, at the Montreal Tandem
accelerator~\cite{Azuelos:2022nbu}, at the JEDI experiment at GANIL,
France~\cite{refId0}, a new experimental setup at LNL,
Legnaro~\cite{Alves:2023ree}, at the nTOF collaboration in
CERN~\cite{Gervino:2023yjm}, and at the Pelletron accelerator in
Melbourne~\cite{Sevior:2023tna}. A detailed list of these experiments
can be found in Ref.~\cite{Krasznahorkay:2025xxy}.

It should be noted that the PADME experiment has recently published a
result of collisions of positron beams on a fixed target. A dedicated
search for the \(X\)-boson has led to the observation of a modest excess
around \SI{16.90}{MeV} with a significance of \reftwo{2.5~\(\sigma\)} (local) and
\reftwo{1.8~\(\sigma\)} (global)~\cite{PADME:2025dla}. On the other hand, no
significant excesses were found in the same mass range in a search 
by the MEG-II experiment~\cite{MEGII:2024urz}. \reftwo{At this point
    this can be attributed to the limited sensitivity of MEG-II and
    should not be interpreted as being in tension with the ATOMKI or the
PADME results}.

The most obvious place to look for the origin of the piling up anomaly
was in nuclear transition form-factors. However, the required form
factor modifications were found to be unrealistic
for\(~^8\)Be~\cite{Zhang:2017zap}. This has led to substantial efforts
at model building leading to the invocation of both pseudoscalar and
vector bosons~\cite{Ellwanger:2016wfe,
DelleRose:2017xil,DelleRose:2018eic} as the candidate \(X\)-boson. In
this work, however, we focus on the protophobic vector boson hypothesis
proposed to explain the anomaly~\cite{Feng:2016ysn}. We describe the
necessary details of this model in the next section.

\section{Model and Parametrisation}%
\begin{figure}[t]
    \centering
    \includegraphics[width=0.8\textwidth]{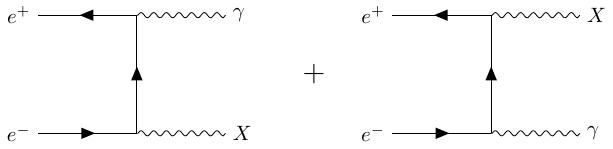}
    \caption{The \(t\)- and \(u\)-channel processes contributing to the
    production of the \(X\)-boson.}
    \label{fig:FD1}
\end{figure}

\label{sec:models} \reftwo{We follow the vectorial model discussed in
    \cite{Feng:2016ysn}. The details of the model can be found in
    \cref{sec:model_cons}. For the purpose of the main text it suffices
    to identify that the \(X\)-boson exclusively decays to an
    electron-positron pair for \(m_X\)=\SI{10}{MeV} to
    \SI{20}{MeV}---the window of our interest. The production of the
    particle is through the t- and u-channel \(e^+e^-\) annihilation, as
shown in \cref{fig:FD1}. We denote the electron coupling of the
\(X\)-boson as \(g^\prime=e\tan\chi\), in terms of which we present our
results. Here, \(e\) is the coupling constant of electromagnetism. }

\begin{figure}[h]
    \centering
    \includegraphics[width=0.48\textwidth]{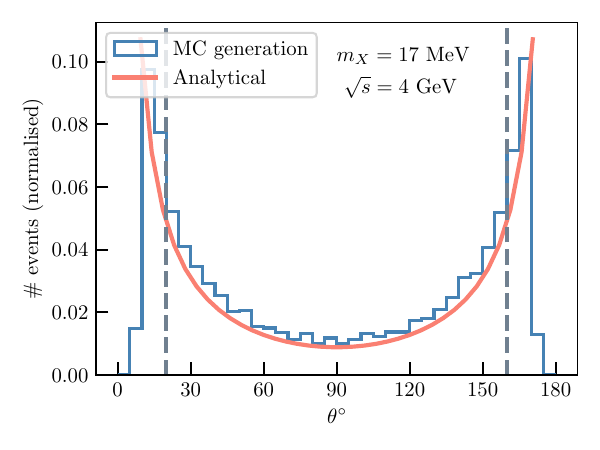}
    \includegraphics[width=0.48\textwidth]{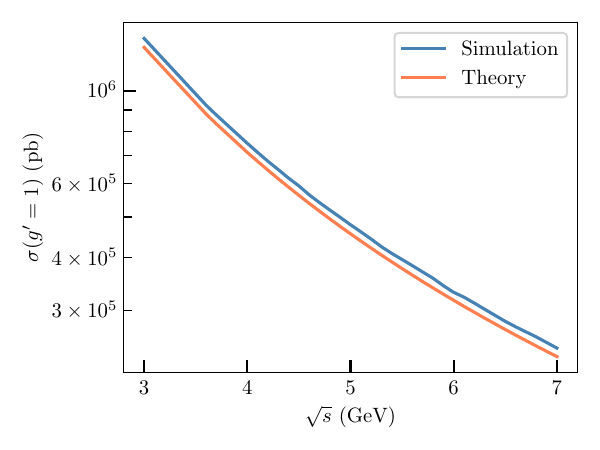}
    \caption{Left: We plot the differential cross section as a function
        of the scattering angle, \(\theta\) for \(m_X=\) \SI{17}{MeV},
        \(\sqrt{s}=\) \SI{4}{GeV}. We plot the cross
        sections generated from MC simulations in \textsf{MG5\_aMC}, and
        calculated from \cref{eq:CS1}. We have indicated \(\theta=20^\circ\) and
        \(\theta=160^\circ\), which are the kinematic endpoints given
        the geometry of the main drift chamber. Right: We
    plot the total cross section as a function of CoM energies. }
    \label{fig:dsec}
\end{figure}

The cross-section of production for \(e^+e^-\to X\gamma\) is given by:
\begin{align}
    \label{eq:CS1}
    \frac{d\sigma}{d\eta}\ &=\ \frac{\alpha_\mathrm{EM}}{2s}
    (g^\prime X_f)^2 
    \frac{\cos^2\vartheta(\eta)(s-m_\chi^2)^2+(s+m_\chi^2)^2}{s(s-m_\chi^2)}\;,
\end{align}
where \(\vartheta(\eta)=2\arctan(e^{-\eta})\), and \(X_f\) is the
\(X\)-boson charge of the fermion. \reftwo{Naturally, there is a
    \(t\)-channel enhancement towards large angles, as can be seen in
    the differential cross-section plot in the left panel
    of~\cref{fig:dsec}, for \(m_X=\) \SI{17}{MeV} and \(\sqrt{s}=\)
    \SI{4}{GeV}. The solid orange line represents our analytical
    calculations. The total cross section is barely sensitive to the
    mass of the \(X\)-boson as the boson mass is much less than the CoM
    energy. However, the
    cross-section varies with the CoM energy, as given in the right
    panel of~\cref{fig:dsec}. In the same figures, we plot the total and
    differential cross-section as obtained from \textsf{MG5\_aMC@NLO} in
blue (further details in \cref{sec:results}). We note that the
analytical and numerical results vary by 4.5\% to 5\%. In our analyses
we consistently use the numerical results.}

In order to expand the scope of the model, we allow the \(X\)-boson to
couple to a generic dark sector as well. We remain agnostic about the
details of the dark sector and do not speculate on the nature of the
couplings, nor do we fix the possible particle content. Instead, we
consider the partial decay width to the dark sector, which we use as a
parameter of the model: 
\begin{align}
    \label{eq:dec}
    \begin{split}
    \Gamma_\mathrm{Tot}\ &=\ \Gamma_\mathrm{VS}+\Gamma_\mathrm{DS}\;, \\
    \mathrm{with},\quad \Gamma_\mathrm{VS}\ &=\  \Gamma_e+\Gamma_\mu+\Gamma_h+\Gamma_\nu+\Gamma_\tau\;,
    \end{split}
\end{align}
where, \(\Gamma_\mathrm{Tot}\) indicates the total decay width.
\(\Gamma_\mathrm{VS}\) and \(\Gamma_\mathrm{DS}\) indicate the partial
widths to the visible and the dark sectors. The visible sector width is
further broken into \(\Gamma_e\), \(\Gamma_\mu\), \(\Gamma_\nu\),
\(\Gamma_\mathrm{h}\), \(\Gamma_\tau\), which are the partial widths to
electrons, muons, hadrons, neutrinos, and taus respectively, \refone{although,
for the region of interest, the visible width is entirely saturated by
decays to \(e^+e^-\) pairs}. For this
analysis, we will work with three benchmark points~\cite{Bandyopadhyay2024}:
\begin{align}
    \label{eq:BP1}
    \Gamma_\mathrm{DS}\ &=\ 0\;;\quad 
    \Gamma_\mathrm{DS}\ =\ 10\Gamma_\mathrm{VS}\;;\quad 
    \Gamma_\mathrm{DS}\ =\ \Gamma_\mathrm{VS}/10\;.
\end{align}

The partial widths are straightforward to compute both above
\(\Lambda_\mathrm{QCD}\) and below the pion mass, for which 
\begin{align}
    \label{eq:BP2}
    \Gamma_{X\to f\bar{f}}\ &=\ \frac{n_c^f}{3} \alpha_{\mathrm{EM}}
    {g^\prime}^2 X_f^2
    m_X \sqrt{1-4\frac{m_f^2}{m_\chi^2}}\left(1+2\frac{m_f^2}{m_X^2}\right)\;,
\end{align}
where, \(n_c^f=1,3\) for leptons and quarks respectively. However, for
\(m_X\sim \Lambda_\mathrm{QCD}\), we have to compute the decay widths to
final state hadrons, using an appropriate effective Lagrangian.
\refone{Detailed discussions on the decay widths can be found
in~\cite{Alexander:2016aln, Ilten:2018crw}.}


The quark-level effective Lagrangian, \reftwo{in the diagonal basis,} that we work with is 
\begin{align}
    \label{eq:efflag}
    \mathcal{L} \supset -\frac{1}{4} X_{\mu\nu}X^{\mu\nu} 
        + \frac{1}{2} m_X^2 X_\mu X^\mu 
        + ig^\prime \sum_f X_f \bar{f}\slashed{X}f 
        + X_\mu\sum_i O_i^\mu 
        + F_{\mu\nu}\sum_iO_i^{\mu\nu}\;.
\end{align}
where \(f\) are the SM fermions, \(X_{f}\) is the \(X\)-boson charge of
\(f\), and \(O_i^\mu\) and \(O_i^{\mu\nu}\) are dimension three and
higher dimensional dark sector operators interacting with the \(X_\mu\)
gauge field and the field-strength tensor, respectively.

Having discussed the production and the decay mechanisms of the
\(X\)-boson, in the next section we discuss the details of the STCF,
especially its drift chamber in order to explain how we intend to
observe the displaced vertices produced by it. 

\section{The STCF main drift chamber}%
\label{sec:STCF}

\begin{table}[htpb]
    \centering
    \caption{CoM energies and corresponding 
    luminosities projected for STCF Phase-I~\cite{Achasov:2023gey}. The 4--7
bucket indicates a 300-point scan with 10 MeV steps, with 
\SI{10}{fb}\textsuperscript{-1} per point. }
    \label{tab:lumi}
    \begin{tabular}{cccccccccccc}
        \toprule
        \(\sqrt{s}\) (GeV) & 3.097 & 3.67 & 3.686 & 3.770 & 4.009 & 4.180 & 4.230 &
        4.360 & 4.420 & 4.630 & 4.0 -- 7.0\\
        \midrule
        \(\mathcal{L}_\mathrm{int}\) (ab\textsuperscript{-1})& 1 & 1 & 1
                                                             & 1
                                                             & 1 & 1 & 1
                                                             & 1 & 1 & 1 & 3  \\
        \bottomrule
    \end{tabular}
\end{table}

The STCF is a proposed \(e^-e^+\) collider with symmetric beam energies
and CoM energies between \SI{2}{GeV} to \SI{7}{GeV}. The peak
instantaneous luminosity is {\(0.5\times 10^{35}\)}
{cm\textsuperscript{-2}s\textsuperscript{-1}} corresponding to a CoM
energy of \SI{4}{GeV} in Phase I. It is on phase I that we exclusively
concentrate in this work. The integrated luminosities corresponding to
the different CoM energies are given in
\cref{tab:lumi}~\cite{Achasov:2023gey, Bao:2025aic}. STCF projects to have hardware and
software upgrades leading to lower backgrounds than current flavor
experiments like Belle-II and LHCb, almost 100\% detection efficiencies,
and almost full detector acceptance~\cite{Achasov:2023gey}. Although
these estimates are \reftwo{optimistic}, they help us in modeling this
initial study. 

\begin{figure}[t]
    \centering
    \includegraphics[width=\textwidth]{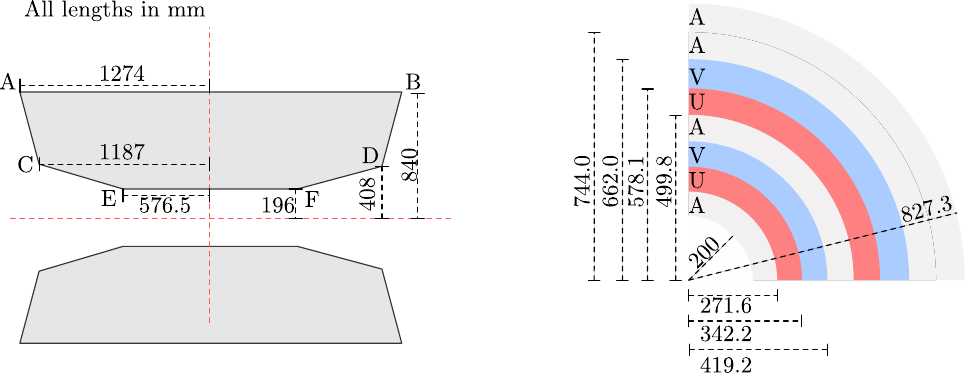}
    \caption{Cross-sectional (right) and longitudinal (left) views of
        the STCF main drift chamber. We show the relevant lengths in
        both the views (all lengths in mm). In the cross-sectional view
        we show the different superlayers in the MDC with both axial (A)
        and stereo (U,V) arrangements. }
    \label{fig:mdc}
\end{figure}

In the barrel region, the STCF detector complex consists of an inner
silicon tracker, surrounded by the MDC, surrounded
by a RICH detector for particle identification, which is enclosed by 
the electromagnetic calorimeter, surrounded by a superconducting magnet
producing \SI{1}{T} magnetic field, with a polar coverage of 93\%. The
outermost layer is a muon counter. Out of these, only the MDC is of
importance to us in this paper. 

\begin{table}[htpb]
    \centering
    \caption{Details of the STCF main drift chamber
    components~\cite{Achasov:2023gey}. `A' represents an axial layer
while `U' and `V' represent stereo layers. }
    \label{tab:super}
    \begin{tabular}{cccccc}
        \toprule
        Superlayer & Radius (mm) & \# layers & Stereo Angle (mrad) & \# cells & Cell Size (mm) \\
        \midrule
        A & 200.0 & 6 & 0 & 128 & 9.8 -- 12.5\\
        U & 271.6 & 6 & 39.3 -- 47.6 & 160 & 10.7 -- 12.9 \\
        V & 342.2 & 6 & \(-\)41.2 -- \(-\)48.4 & 192 & 11.2 -- 13.2 \\
        A & 419.2 & 6 & 0 & 224 & 11.7 -- 13.5\\
        U & 499.8 & 6 & 50.0 -- 56.4 & 256 & 12.3 -- 13.8 \\
        V & 578.1 & 6 & \(-\)51.3 -- \(-\)57.2 & 288 & 12.6 -- 14.0 \\
        A & 662.0 & 6 & 0 & 320 & 13.0 -- 14.3\\
        A & 744.0 & 6 & 0 & 352 & 13.3 -- 14.5\\
        \midrule
        Total & 200 -- 827.3 & 48 & & 11520 &\\ 
        \bottomrule
    \end{tabular}
\end{table}

In \cref{fig:mdc} we have drawn cartoons of the cross-section and the
longitudinal-section of the STCF MDC, as obtained from their Conceptual
Design Report (CDR) on detectors~\cite{Achasov:2023gey}. The MDC extends
in the radial direction from \SI{20}{cm} to \SI{83}{cm}, with a total of
48 layers of drift cells. The drift cells are arranged in 8 superlayers,
the details of which are given in \cref{tab:super}. The cell dimensions
are designed to range from 0.98 -- \SI{1.25}{cm} in the innermost
superlayer to 1.33 -- \SI{1.45}{cm} in the outermost superlayer.

The MDC is expected to i) measure momentum with an excellent resolution
of \(<0.5\%\) at \SI{1}{GeV}, ii) reconstruct charged tracks in tandem
with the inner trackers, and iii) measure the energy loss of the
particles (\(dE/dx\)) for PID\@. The track reconstruction methods and
algorithms planned for STCF have been rigorously studied in Refs.
\cite{Ai:2024yqx,Ai:2024hnv}. In this work we rely on the tracking
efficiency of the MDC exclusively, and do not take into account the
inner trackers as the inner trackers are too close to the IP, and SM
displaced vertex processes dominate. We model the STCF MDC detector
geometry using the \textsf{TrackEff} package~\cite{Bertholet:2025lcr},
and it is with the help of \textsf{TrackEff} that we obtain the vertex
reconstruction efficiency (VRE). The 48 layers of the MDC are grouped 
into  superlayers, each containing 6 layers. The number of drift cells
in the \(r\)-\(\phi\) plane increases from 128 in the innermost layer to
352 in the outermost layer. The drift cells are approximately squares in
cross-section, with the size gradually increasing within each
superlayer. This morphology was implemented in \textsf{TrackEff}.

Using the momentum and the position of the decay vertex of the
\(X\)-boson, obtained from our simulations, \textsf{TrackEff}
reconstructs the trajectories of the daughter electrons by propagating
them through the uniform magnetic field, of \SI{1}{T}, inside the
fiducial volume of the detector. The path of each trajectory in each
drift cell is divided into \(n_s\) segments of length \(\Delta_s\). In
our analysis we assume the default length size of
\(\Delta_s\)=\SI{1}{mm}. The minimum \(n_s\) required within a drift
cell of each layer can be adjusted to control the hit acceptance
criteria. To account for the increasing cell size within each
superlayer, we modified the TrackEff configuration to reflect the MDC
layer arrangement. We set this threshold to \(n_s^\mathrm{inner}=5\) for
cells in the innermost superlayer, while for the remaining superlayers
we use \(n_s=10\). This choice ensures that smaller cells near the
interaction point, more sensitive to highly boosted tracks, require
fewer steps, whereas the larger cells at higher radii require more steps
to maintain consistent hit acceptance. For a track to be considered
reconstructed, we require it to have produced hits in at least 24
detector cells. \reftwo{Having discussed all the required
pieces that go into our analysis, in the next section we give
the actual analysis details and our findings.}

\section{Analyses and Results}%
\label{sec:results}

\begin{figure}[ht]
    \centering
    \includegraphics[scale=0.87]{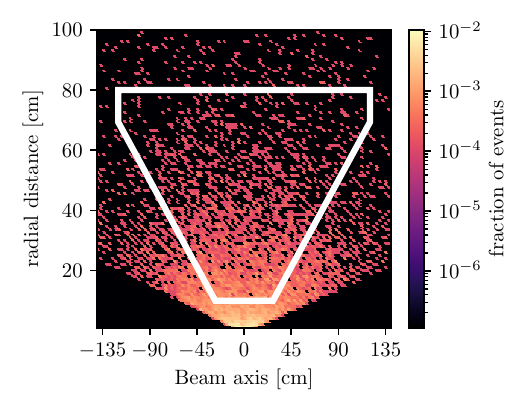}
    \caption{Heatmaps showing the distribution of the decay vertex of
        the \(X\)-boson. We have plotted the outline of the region of
        acceptance of vertices in white. The plot shows 
        events for \(\sqrt{s}=\)\SI{4}{GeV}. The
    parameter point considered is \(m_X=\) \SI{17}{MeV} and
\(g^\prime\sim 10^{-5}\), which translates to \({\beta\gamma c
\tau}\simeq\) \SI{49}{cm}.} \label{fig:heat}
\end{figure}

For our analyses we generated \(10^4\) events per mass point per CoM
energy in
\textsf{MG5\_aMC@NLO}\footnote{\href{https://launchpad.net/mg5amcnlo}{https://launchpad.net/mg5amcnlo},
Open Source}~\cite{Alwall:2014hca}. Instead of a 300-point scan of the 4
-- \SI{7}{GeV} range with 10 fb\(^{-1}\) per point, \reftwo{as planned
by the collaboration} (\cref{tab:lumi}), we perform a 30-point
scan with 100 fb\(^{-1}\) per point. The total number of events we work
with per mass-coupling point is then \(4\times 10^5\), corresponding to
the CoM energies given in \cref{tab:lumi}. The expected
number of DV events per CoM energy, as computed from the
cross-section is then: 
\begin{align}
    \label{eq:lumi}
    \begin{split}
        N^i(m_X,g^\prime, \Gamma_\mathrm{DS}) \ &=\ \mathcal{L}_\mathrm{int}^i\times
        \sigma^i(m_X,g^\prime)\times \mathrm{BR}(m_X, \Gamma_{DS}) \times \mathrm{eff}^i(m_X,g^\prime, \Gamma_\mathrm{DS})\;;\\
        \mathrm{with,}\quad N_\mathrm{tot}\ &=\ \sum_i N^i\;,
    \end{split}
\end{align}
\reftwo{The superscript \(i\) indicates the sample index for different
\(\sqrt{s}\) points given in \cref{tab:lumi}}. Here,
\(N^i\) is the number of events, \(\mathcal{L}_\mathrm{int}^i\) is the
integrated-luminosity, \(\sigma^i\) is the cross-section of \(e^+e^-\to
\gamma X\), \(\mathrm{BR}\) is the branching ratio of the \(X\)-boson to
two electrons, and \(\mathrm{eff}^i\) is the overall detection
efficiency of the displaced vertex.  The total number of events,
over all CoM energies, is \(N_\mathrm{tot}=\sum_i N_i\).

The generated events are binned by scattering angle, in bins of
\(1^\circ\). The \(X\)-bosons produced in each bin are made to decay at a
random distance away from the IP, \reftwo{as per its lifetime and
momentum}. 

\begin{table}[htpb]
    \centering
    \caption{List of the cuts that we impose for the event selection. The
    \(p_T^\ell\) cut is on the electrons and positrons which are produced
from the \(X\)-boson decay. The geometric cuts are all on the position
of the displaced vertex.}
    \label{tab:cut}
    \begin{tabular}{c c c c c c c}
        \toprule
        \(p_T^\ell\) & \(l_\mathrm{min}\) & \(l_\mathrm{max}\) &
        \(z_\mathrm{min}\) & \(z_\mathrm{max}\) &
        \(\theta_\mathrm{min}\) & \(\theta_\mathrm{max}\)\\
        \midrule
        \SI{0.1}{GeV} & \SI{10}{cm} & \SI{80}{cm} & $-$\SI{120}{cm} &
        \SI{120}{cm} & \(30^\circ\) & \(150^\circ\)\\
        \bottomrule
    \end{tabular}
\end{table}

We apply a series of analysis-level cuts on the position of the decay
vertices of the produced \(X\)-boson and on the momenta of the produced
\(e^+\) and \(e^-\). These cuts have been noted in \cref{tab:cut}. The
cut on the momenta of the daughter \(e^+\) and \(e^-\) has been imposed
for efficient track reconstruction. In \cref{sec:app}, we demonstrate
the effect of varying this cut on discovery potential. The lower limit
on the distance, \(l_\mathrm{min}\), ensures that we are far away from
the IP for long-lived SM backgrounds from \(K_S^0\to\pi^+\pi^-\),
\(\Lambda\to p\pi^-\) to drop
off~\cite{Bertholet:2021hjl,Bandyopadhyay:2022klg}. The upper limit,
\(l_\mathrm{max}\), is determined by the extent of the detector. As we
pass the events through another layer of selection through
\textsf{TrackEff}, which checks if the event translates to
reconstructible tracks, we keep the upper limit aggressively close to
the detector endpoint at this level.

In \cref{fig:heat} we plot the normalised distribution of the spatial
location of the decay vertices of the \(X\)-boson for the \(10^4\)
events we generate for \(\sqrt{s}=\) \SI{4}{GeV}, \(m_X\sim\)
\SI{17}{MeV}, and \(g^\prime=10^{-5}\). This translates to \(\beta\gamma
c\tau\simeq\) \SI{49}{cm}\footnote{As for two body decay
\(|\vec{p}|=\frac{s-m^2}{2\sqrt{s}}\).}. We have binned the vertices on
a 100\(\times\)100 grid, from \SI{-140}{cm} to \SI{140}{cm} in the \(z\)
direction and from \SI{0}{cm} to \SI{100}{cm} in the radial direction.

It is clear that most of the \(X\)-bosons still decay near the IP,
despite a large characteristic decay length. At generation level we have
imposed a cut on rapidity of \(|\eta|=2.5\) (\(\theta\sim10^\circ\)),
hence, there are no events in the corresponding region in
\cref{fig:heat}.
\begin{figure}[ht]
    \centering
    \includegraphics[scale=1]{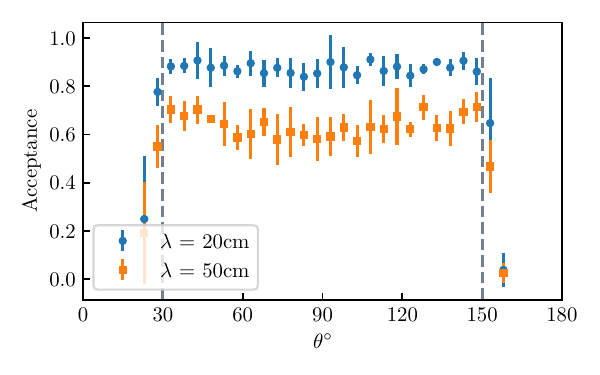}
    \caption{\reftwo{Vertex reconstruction efficiency distribution
    as a function of the scattering angle, binned in \(5^\circ\) bins.
    The efficiency drops drastically for large angles.}}
    \label{fig:vre}
\end{figure}

The \(e^+e^-\) tracks produced at the site of the \(X\)-boson decay
vertex are passed through \textsf{TrackEff} to see which tracks are
accepted as per the selection criteria discussed in the last section.
After performing the analysis over \(4\times 10^5\) events (after
combining all the \(\sqrt{s}\)) for each parameter point, we multiply
the number of accepted events by the relevant efficiency factor. Note,
the number of accepted events indicate the number of events
reconstructed after passing through our simulation chain. An event is
considered accepted when both the tracks corresponding to the event are
reconstructed. The acceptance constitutes the detector efficiency in
\cref{eq:lumi}. It is defined as:
\begin{align}
    \label{eq:acc}
    \epsilon^i\ \equiv\ \mathrm{eff}^i\ &=\ \frac{N_\mathrm{acc}^i}{N_\mathrm{MC}^i}\;,
\end{align}
where \(N_\mathrm{MC}\) is the total number of MC generated events which
pass the cuts given in \cref{tab:cut}, and \(N_\mathrm{acc}\) are the
events out of it which are reconstructed by \textsf{TrackEff}.  

\begin{figure}[t]
    \centering
    \includegraphics[scale=0.85]{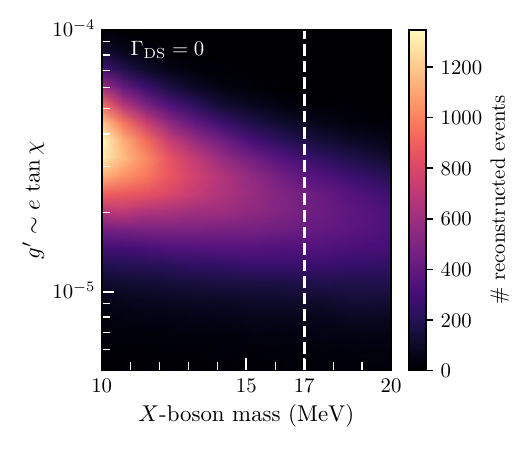}
    \includegraphics[scale=0.85]{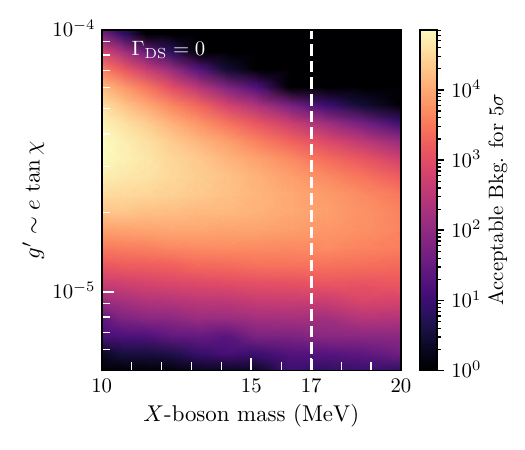}
    \caption{On the left panel we plot the number of signal events after
        detector simulation via \textsf{TrackEff}, as a function of the
        mass of the \(X\)-boson and its coupling strength. On the right panel,
        we plot the number of background events that can be tolerated for
        \(\reftwo{5~\sigma}\) discovery of the \(X\)-boson, as a function of the
        same mass and couplings.}
    \label{fig:fin}
\end{figure}

In \cref{fig:vre} we plot eff\textsuperscript{i} \reftwo{as a function of
\(\theta\) for \(m_X=\) \SI{17}{MeV} and} \(10^4\) events corresponding to \(\beta\gamma c\tau\)=
\SI{10}{cm} and \SI{50}{cm}.  We note, as the decay length increases, the plot
becomes more erratic. This is expected as the number of detector layers
available to a track generated far away from the IP will be much less.
Hence, the transversality of the track will play an important role.
Tracks which are more longitudinal will have access to more cells to hit
than tracks which are more transverse. \reftwo{This is also visible
    within the plot for a particular decay length. The uncertainty in
    acceptance decreases as we go further from the central region.
    The probability of tracks being detected stabilises for longitudinal
tracks. Also, the shape of the differential distribution,
\cref{fig:dsec}, imposes that fewer events travel in the transverse
direction than the longitudinal direction leading to less statistics and
more uncertainty.} Also, note that the acceptance
falls drastically at large and small angles. This is due to the detector
geometry, as discussed in the last section. \reftwo{The kinematic
end-points are given by
\(20^\circ<\theta<160^\circ\). However, the actual falling off
starts about \(10^\circ\) before the kinematic end points. We note that
as the \(X\) boson track becomes tangential to the detector edges, it
becomes difficult for both of the ensuing electrons to be reconstructed.
In most cases, the opening angle would make one lepton to go inside the
detector and the other lepton to go out of it. Hence, at analysis level
we impose the cuts: \(30^\circ<\theta<150^\circ\). }


\begin{figure}[t]
    \centering
    \includegraphics[scale=0.85]{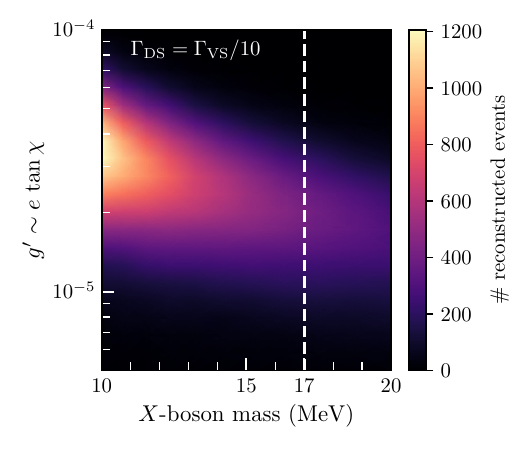}
    \includegraphics[scale=0.85]{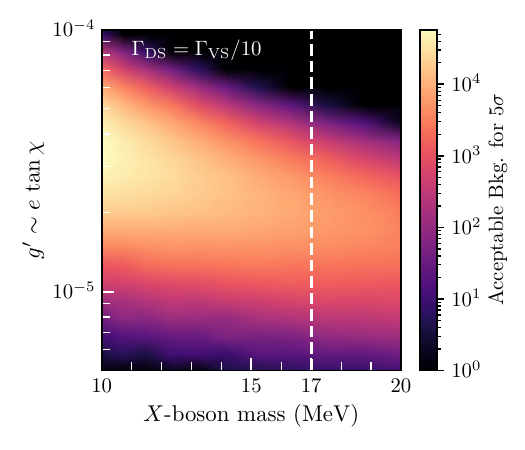}
    \includegraphics[scale=0.85]{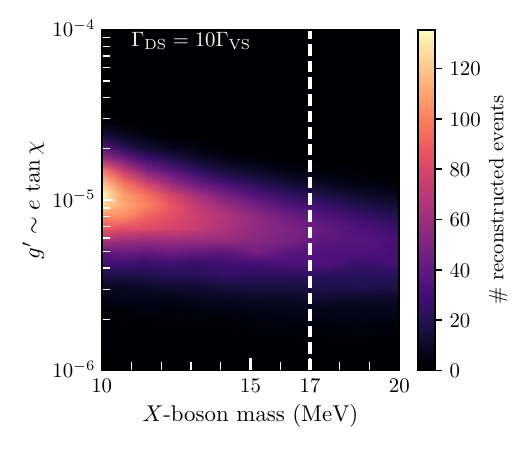}
    \includegraphics[scale=0.85]{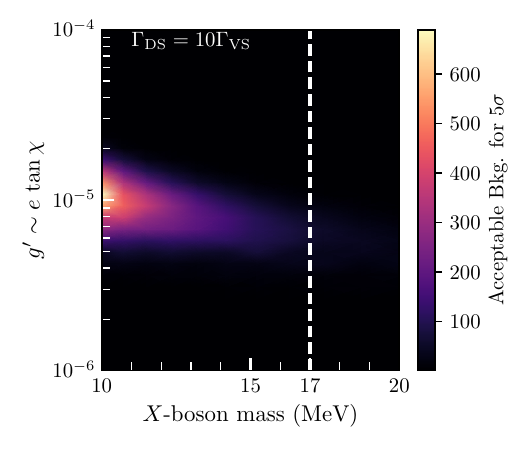}
    \caption{On the left panel we plot the number of signal events after
        detector simulation via \textsf{TrackEff}, as a function of the
        mass of the \(X\)-boson and its coupling strength. On the right
        panel, we plot the number of background events that can be
        tolerated for \(\reftwo{5~\sigma}\) discovery of the
        \(X\)-boson, as a function of the same mass and couplings. The
    top row is for the \(\Gamma_\mathrm{DS}=\Gamma_\mathrm{VS}/10\) and
the bottom row is for the \(\Gamma_\mathrm{DS}=10\Gamma_\mathrm{VS}\).}
    \label{fig:fin_mod}
\end{figure}

Although we consider displaced vertices at least \SI{10}{cm} from
the IP to mitigate SM background, we do not get a background-free
environment. There will still be SM displaced tracks due
to \(K_S^0\to\pi^+\pi^-\) and \(\Lambda\to
p\pi^-\)~\cite{Bertholet:2025lar} decays. Also, there will be
backgrounds from pair production by photons interacting with the
detector material~\cite{Ferber:2022ewf,Jaeckel:2023huy}. In addition,
hadronic interactions with the detector material will produce mesons,
protons, and nuclear fragments which will give rise to tracks. There
will also be backgrounds from \reftwo{accidental and coincidental} crossing of
tracks~\cite{Bertholet:2025lar}. It is beyond the scope of this paper to
perform the necessary detector simulations to estimate the exact
fraction of background events that survive the cuts, as backgrounds for
displaced vertices necessarily requires a simulation of interaction of
radiation with matter. To compensate for this, we keep the number of
background events as a parameter of our search. Instead of presenting a
deterministic \(\reftwo{5~\sigma}\) contour of discovery, we find out the number
of background events which can be tolerated for \reftwo{\(5~\sigma\)} discovery,
per mass and coupling point. 

The median discovery significance, with \(n_s\) number of signal events
and \(n_b\) number of background events is given
by~\cite{Li:1983fv,Cousins:2007yta,Cowan:2010js,Kumar:2015tna}:
\begin{align}
    \label{eq:disc}
    Z_\mathrm{disc}\ &=\ \reftwo{\sqrt{2\left[(n_s+n_b)\ln(1+n_s/n_b)-n_s\right]}}\;.
\end{align}
We solve \cref{eq:disc} as a function of \(n_b\) with
\(Z_\mathrm{disc}=5\), to find the number of background events
admissible. The signal events, \(n_s\), are what we find in our
simulations, \emph{i.e.}, as given in \cref{eq:lumi}.  

In the left panel of \cref{fig:fin}, we plot the number of true events
corresponding to the mass and the coupling of the \(X\)-boson, as
obtained from \cref{eq:lumi} and \cref{eq:acc}. In the right panel
of~\cref{fig:fin}, we plot the number of background events that can be
tolerated to produce \reftwo{\(5~\sigma\)} discovery significance, as dictated by
\cref{eq:disc}. From \cref{fig:fin}, it is clear that for extended
regions of the parameter space around \SI{17}{MeV}, the \(X\)-boson
search can tolerate up to \(10^4\) background events for \reftwo{\(5~\sigma\)}
discovery. \reftwo{Note that, we have simulated over a matrix of 60 logarithmically
    distributed bins between \(10^{3}\) and \(10^{-7}\) for \(g^\prime\)
    and 10 linearly distributed bins for \(m_X\) between 10 and 20 MeV.
    For the final results we did a 2-dimensional interpolation for the
parameter space.} 

We repeat the same exercise for the two other branchings given in
\cref{eq:BP1}. For \(\Gamma_\mathrm{DS}=\Gamma_\mathrm{VS}/10\), we do not
expect much difference from the \(\Gamma_\mathrm{DS}=0\) case.  
However, for \(\Gamma_\mathrm{DS}=10\Gamma_\mathrm{VS}\), the situation
is dramatically different, as for discovery, the number of backgrounds
need to be extremely low---a few hundreds of events.  

\begin{figure}[htpb]
    \centering
    \includegraphics[scale=1]{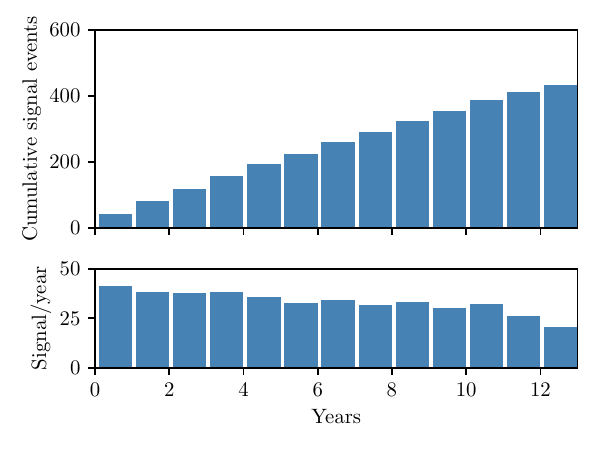}
    \caption{The number of signal events to be generated by STCF per
        year (bottom plot) and the cumulative signal events collected
        after each year (top plot). On the top plot, the vertical lines
        denote the number of years required to signal a discovery for
        different numbers of background events. The values of the mass
    and the coupling are respectively \(m_X=\)\SI{17.5}{MeV} and
\(g^\prime\sim 2\times 10^{-5}\).}
    \label{fig:years}
\end{figure}

\reftwot{To get a rough estimate of the time that should be required to
    potentially detect an \(X\)-boson, we plot the number of signal
    events that would be collected per year by STCF in the bottom panel
    of \cref{fig:years}. Note that, as the cross section falls with
    increasing CoM, \cref{fig:dsec}, the number of events decrease with
    each passing year, as STCF is expected to increase the CoM energy
    each year, \cref{tab:lumi}. In the top panel of the same figure, we plot
    the cumulative number of events collected after every year. For the
    plot we use \(m_X=\)\SI{17.5}{MeV} and \(g^\prime\sim 2\times
    10^{-5}\). The acceptable background for a \(5 \sigma\) discovery
after year \(n_\mathrm{years}\), can be extracted from the plot using
\cref{eq:disc}. Of course, the accuracy and precision of the extraction
will be correct up to our simplifying assumption of no systematics.}

\section{Conclusion}%
\label{sec:conclusion}
The anomalies in nuclear transitions reported by the ATOMKI
collaboration  provide us with overwhelming hints of physics beyond the
Standard Model. The protophobic model is a prime candidate that
ameliorates the tension between theory expectations and experimental
observations. However, some negative results, like the one reported by
MEG-II~\cite{MEGII:2024urz}, and tensions with electroweak precision
observables~\cite{DiLuzio:2025ojt} implore us to keep looking for the
protophobic boson at other experimental facilities. In this work, we
initiate this effort for the proposed Super \(\tau\)-Charm Facility
(STCF). We observe that a protophobic boson of mass around \SI{17}{MeV}
can be discovered at \reftwo{5~\(\sigma\)} C.L. by the STCF for
background events as large as \(10^4\). In essence, this work gives
another use case to the already rich STCF physics programme. \reftwo{Our results
are encouraging enough for the members of the STCF collaboration to
address precisely the feasibility for the observation of X17 with more
detailed and dedicated simulation based on programs like GEANT-4}. 

\medskip
\textbf{Acknowledgements:} 
    We thank Jim Libby, Biplob Bhattacherjee, Anindya Datta, and Gautam
    Bhattacharyya for useful discussions and suggestions. Other than the
    packages mentioned in the text, all computations and visualisations
    have been done using Python 3.13.7 and its scientific stack of
    Numpy~\cite{harris2020array}, SciPy~\cite{2020scipy-nmeth},and
    Matplotlib~\cite{Hunter:2007}. 

\appendix
\crefalias{section}{appendix}
\section{\texorpdfstring{Results for varying \(p_T\) cuts}{Results for
        varying pT cuts}}
\label{sec:app}
The analyses in our main text is crucially based on the STCF claim of
targeting near 100\% detection efficiencies~\cite{Achasov:2023gey,
Ai:2023ukc, Ai:2024hnv} of
charged tracks above \(p_T^\ell\)=\SI{0.3}{GeV} and about 90\%
efficiency \reftwo{down to} \SI{0.1}{GeV}. As a result, we
imposed a mild \(p_T^\ell\) cut on the charged tracks of \SI{0.1}{GeV}. 
In this appendix we explore the effects of imposing stronger
\(p_T^\ell\) cuts on the final state leptons. This is a contingency
study in case STCF fails to reach the \reftwo{design goals}. 

\begin{figure}[h]
    \centering
    \includegraphics[width=0.45\textwidth]{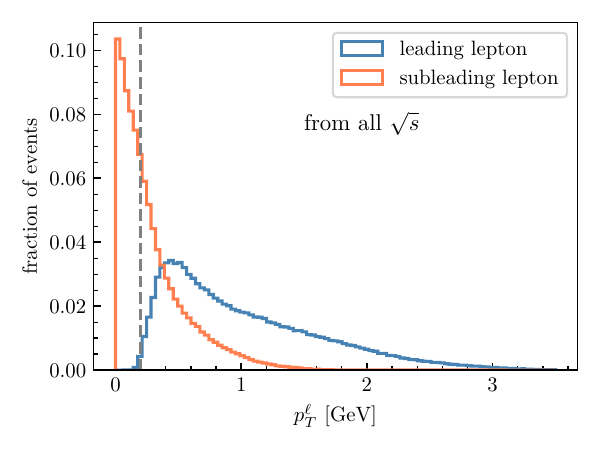}
    \includegraphics[width=0.45\textwidth]{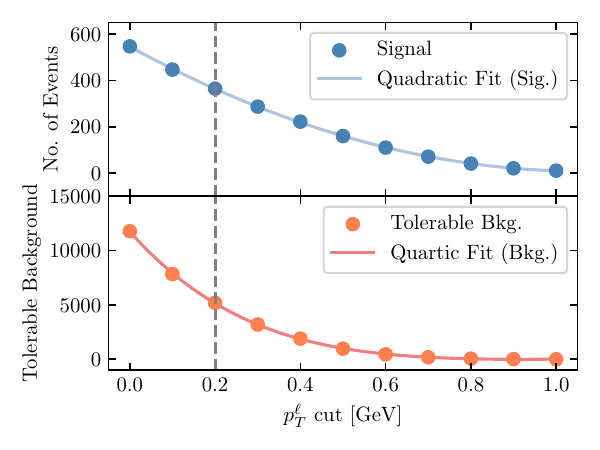}
    \caption{In the left panel we \reftwo{show the distribution of} the leading and subleading leptons
    according to their transverse momentum. In red we show the
number of subleading leptons per transverse momentum bin and in blue we show the number of
leading leptons per bin. In the right panel we plot the number of
accepted signal events for different \(p_T^\ell\) cuts (blue). With an increasing
cut value, the number of signal events drop off quadratically. We also
plot the corresponding tolerable background (red) for
\reftwo{\(5~\sigma\)} discovery.
We find that it drops off as a \reftwo{quartic}. }
    \label{fig:subleading}
\end{figure}

In the left panel of \cref{fig:subleading}, we plot the \(p_T\) of the leading and
subleading leptons for all the events we generated across the different
\(\sqrt{s}\). The figure makes it clear that a large fraction of
subleading leptons have \(p_T\) less than \SI{0.2}{GeV}. Here, we focus
on \SI{0.2}{GeV} as BESIII can efficiently track charged particles till
that threshold~\cite{Jia:2010zz,Jia:2024rbx, BESIIIslides1}. 

In the right panel of the figure, we choose a mass-coupling point,
\SI{17.5}{MeV} -- \(2\times 10^{-5}\), and compute the number of accepted
events for different \(p_T^\ell\) cuts ranging from
\(p_T^\ell\)=\SI{0}{GeV} to \(p_T^\ell\)=\SI{1}{GeV}. We find that the
number of signal events fall off quadratically with an increasing
\(p_T^\ell\) cut. We also plot the corresponding background events that
can be tolerated for \reftwo{\(5~\sigma\)} discovery and find that it falls
as a \reftwo{quartic} with an increasing \(p_T^\ell\) cut. What the figure tells us
is that for a \(p_T^\ell\) cut between \SI{0.1}{GeV} and \SI{0.2}{GeV},
we can tolerate between \(\sim\) 9000 to \(\sim\) 5000 background events
for a \reftwo{\(5~\sigma\)} discovery. For calculating the number of signal
events, we have used \(\Gamma_\mathrm{DS}=0\). 

\section{Model Considerations}
\label{sec:model_cons}
For this work, we have adopted the model proposed in Ref.
\cite{Feng:2016ysn}. We consider the protophobic boson to be the gauge
boson associated with the \(U(1)_B\) of baryon number. Therefore, it
couples to fermions with a charge equal to baryon number. The \(U(1)_B\)
symmetry is spontaneously broken to identity by a Higgs mechanism,
generating mass for the \(X\)-boson. The \(U(1)_B\) mixes via the
kinetic terms~\cite{Okun:1982xi,Galison:1983pa,Holdom:1985ag,Fayet:1990wx,Curtin:2014cca} with the
\(U(1)_\mathrm{EM}\), with the Lagrangian:
\begin{align}
    \label{eq:lag1}
    \mathcal{L}\ &\supset\ -\frac{1}{4}
        \widetilde{F}^{\mu\nu}\widetilde{F}_{\mu\nu}
    -\frac{1}{4}
        \widetilde{X}^{\mu\nu}\widetilde{X}_{\mu\nu}
    -\frac{\sin\chi}{2}
        \widetilde{F}^{\mu\nu}\widetilde{X}_{\mu\nu}
        + \frac{1}{2} (D_\mu\phi)^\dagger D^\mu\phi
        + \sum_f \bar{f} i\slashed{D} f\;.
\end{align}
Here, \(\widetilde{F}_{\mu\nu}\) and \(\widetilde{X}_{\mu\nu}\) are the
field-strength tensors corresponding to the \(U(1)_\mathrm{EM}\) field
and the \(U(1)_B\) field respectively. The sum runs over all fermions
\(f\) and we have 
\begin{align}
    \label{eq:lag2}
    D_\mu \phi\ &=\ \left(\partial_\mu + i e\epsilon_BB_\phi
        \widetilde{X}_\mu\right)\phi\;,\\
    D_\mu f\ &=\ \left(\partial_\mu + ie Q_f\widetilde{A}_\mu + ie\epsilon_B
        B_f\widetilde{X}_\mu\right)f\;.
\end{align}
Here, \(\widetilde{A}_\mu\) and \(\widetilde{X}_\mu\) are the fields
corresponding to the photon and the \(U(1)_B\) respectively, \(Q_f\) and
\(B_f\) are the electric charge and the baryon number of the fermion in
question, and \(B_\phi\) is the \(X\)-charge of the scalar. Finally,
\(e\epsilon_B\) is the coupling strength associated with the
\(\tilde{X}\)-boson.

To go to the canonically diagonal kinetic basis, we diagonalise the kinetic
sector matrix via the triangular matrix
\begin{align}
    \label{eq:lag3}
    \begin{pmatrix} \widetilde{A}^\mu \\ \widetilde{X}^\mu \end{pmatrix}
        \ &=\ 
            \begin{pmatrix} 
                1 & -\tan\chi \\ 
                0 & \sec\chi 
            \end{pmatrix}
            \begin{pmatrix} A^\mu \\ X^\mu \end{pmatrix} 
            \;,
\end{align}
where \(A^\mu\) and \(X^\mu\) are the fields in the canonically diagonal
basis. In this basis, the covariant derivatives become
\begin{subequations}
\begin{align}
    \label{eq:charge_def}
    D_\mu \phi\ &=\ \left(\partial_\mu + i e\epsilon_B\sec\chi B_\phi{X}_\mu\right)\phi\;,\\
    D_\mu f\ &=\ \left(\partial_\mu + ie Q_f{A}_\mu 
        + ie(-Q_f\tan\chi + \epsilon_B \sec\chi B_f ){X}_\mu\right)f\;.
\end{align}
\end{subequations}
\begin{table}[htpb]
    \centering
    \caption{The \refone{\(U(1)_X\)} charges of the fermions of one generation in units of
    \(e\tan\chi\).}
    \label{tab:charge}
    \begin{tabular}{lllll}
        \toprule
        Fermion & \(q_u\) & \(q_d\) & \(e\) & \(\nu\) \\
        \midrule
        Charges & \(-\frac{1}{3}\) & \(\frac{2}{3}\) & 1 & 0\\
        \bottomrule
    \end{tabular}
\end{table}
Now, imposing, \(\epsilon_B= \sin\chi\),  we get the \(X\) charges of the
fermions, as given in \cref{tab:charge}. \refone{Note that, setting
\(\epsilon_B= \sin\chi\) is tantamount to a fine tuning. However,
as the kinetic mixing parameter is in any case a free parameter and does
need an initial value to determine its running, our choice is as good as
any other choice for it. In essence, this choice defines a particular
instance of the model. Other choices will lead to other instances.}

\refone{This choice ensures that the up quark has charge \(-1/3\) and the
down quark has charge \(2/3\)---making the proton free of any \(X\)
charge. Hence, the name `protophobic'. This demand of protophobia
automatically ensures that the electron picks up a unit charge under the
\(X\) boson, while the neutrino remains neutral~\cite{Feng:2016ysn}.}

The corresponding covariant derivatives are:
\begin{subequations}
\begin{align}
    \label{eq:charge_def_1}
    D_\mu \phi\ &=\ \left(\partial_\mu + i e\tan\chi B_\phi X_\mu\right)\phi\;,\\
    D_\mu f\ &=\ \left(\partial_\mu + ie Q_f A_\mu + ie\tan\chi X_f{X}_\mu\right)f\;.
\end{align}
\end{subequations}
Then \(g^\prime=e\tan\chi\) is the effective coupling strength of the
\(X\)-boson, and \(X_f=B_f-Q_f\) is the effective charge of the fermion
\(f\). Now, once the scalar \(\phi\) acquires a vev
\(\phi\to\phi+v_\phi\), the gauge boson \(X_\mu\) acquires a mass,
\(m^2_\chi=e^2\tan^2\chi B_\phi^2 v_\phi^2\). For the couplings we are
looking at, \(g^\prime\sim 10^{-3}\) -- \(10^{-6}\) GeV, the singlet
scalar ends up in the multi TeV ranges, and is safe from constraints.
Note that, we are writing the charges much below the electroweak
symmetry breaking scale. Also, \(\epsilon_B=\sin\chi\) can only be
imposed at any one scale in the evolution of the coupling. The two
entities will renormalise differently as the scale runs. However, we can
always set the equality at \(\sim\) \SI{17}{MeV} so that we don't have
to worry about running effect in the narrow window around \(\sim\)
\SI{17}{MeV}, where we concentrate. 

Also note, the \(U(1)_B\) symmetry is anomalous as the gauge currents do
not cancel either mixed-gauge or gauge-gravity anomalies with the SM
particle content. However, this has to be viewed in context. As we are
performing a low-energy study, our theory can always be viewed as the
remnant of a UV-complete theory where anomaly cancellation is ensured by
additional chiral (under \(U(1)_B\)) fermions above the electroweak
scale, in the far UV (check, e.g., Ref.~\cite{Feng:2016ysn}).
\refone{Among other effects, the heavy fermions might induce kinetic
    mixing, however, we have already included that in our setup. 
} 

\refone{UV completions of gauged \(U(1)_B\) have been proposed in Refs.
    \cite{Duerr:2013dza,Duerr:2014wra,Fornal:2015boa}. The completions
    involve the addition of colour singlet fermions which are vectorial
    under the SM gauge group but chiral under \(U(1)_B\). While allowing
    the corresponding gauge boson to be \(\sim 17\) MeV, the vector-like
    singlets can be of TeV order, getting mass from
    \(v_\phi\)---regions where LHC searches are not sensitive yet (see,
    e.g., Refs. \cite{CMS:2022nty,ATLAS:2024mrr,CMS:2025urb}). For other
    constraints on this completion check Ref~\cite{Feng:2016ysn}. We
    stress that this is only the simplest UV completion. There is no
    obvious roadblock for creating other completions where the masses of
    the heavy fermions are decoupled from the mass of the light gauge
    boson, leading to the possibility of the fermions being
super-heavy---far above the TeV scale.}  

\printbibliography[type=article]
\end{document}